\documentclass[10pt]{article}
\usepackage[textwidth=15cm, textheight=22cm]{geometry}

\usepackage{amsfonts}
\usepackage{amsmath}
\usepackage{amssymb}
\usepackage{amsthm}

\usepackage{nicefrac}

\usepackage{multicol}

\usepackage[dvipsnames]{xcolor}
\definecolor{myblue1}{RGB}{241,238,246}
\definecolor{myblue2}{RGB}{189,201,225}
\definecolor{myblue3}{RGB}{116,169,207}

\usepackage{authblk}
\usepackage{graphicx}
\usepackage{epic}
\usepackage{overpic}
\usepackage{subfig}

\usepackage{tikz}
\usetikzlibrary{arrows.meta,shapes,calc}
\usetikzlibrary{positioning,shapes,arrows}

\usepackage{epstopdf}

\usepackage{todonotes}

\usepackage{url}

\newcommand{\AAA}{\mathcal{A}}

\newcommand{\NN}{\mathcal{N}}

\newcommand{\UU}{\mathcal{U}}

\newcommand{\N}{\mathbb{N}}

\newcommand{\R}{\mathbb{R}}

\newcommand{\dx}{\dot{x}}

\begin{document}

\title{Uniting Parametric Uncertainty and Tipping Diagrams}
\author[1]{Kerstin Lux}
\author[2]{Peter Ashwin}
\author[3]{Richard Wood}
\author[1,4]{Christian Kuehn}
\affil[1]{Technical University of Munich, Department of Mathematics, Boltzmannstr.~3, 85748 Garching, Germany (kerstin.lux@tum.de)}
\affil[2]{Department of Mathematics, University of Exeter, Exeter EX4 4QF, UK}
\affil[3]{Met Office Hadley Centre, Exeter, UK}
\affil[4]{Complexity Science Hub Vienna, Josefst\"adter Str.~39, 1080 Vienna, Austria}

\maketitle

\begin{abstract}
Various subsystems of the Earth system may undergo critical transitions by passing a so-called tipping point, under sustained changes to forcing. For example, the Atlantic Meridional Overturning Circulation (AMOC) is of particular importance for North Atlantic heat transport and is thought to be potentially at risk of tipping. Given a model of such a subsystem that accurately includes the relevant physical processes, whether tipping occurs or not, will depend on model parameters that typically are uncertain. Reducing this parametric uncertainty is important to understand the likelihood of tipping behavior being present in the system and possible tipping locations. In this letter, we develop improved estimates for the parametric uncertainty
by inferring probability distributions for the model parameters based on physical constraints and by using a Bayesian inversion technique. To visualize the impact of parametric uncertainty, we extend classical tipping diagrams by visualizing probabilistic bifurcation curves according to the inferred distribution of the model parameter. Furthermore, we highlight the uncertain locations of tipping points  along the probabilistic bifurcation curves.
We showcase our probabilistic visualizations of the tipping behavior using a simple box-model of the AMOC, the Stommel-Cessi model \cite{Cessi.1994}.
\end{abstract}


\twocolumn
	
\section{Introduction}

Several subsystems or ``elements'' of the Earth system have been identified as potentially at risk of tipping \cite{Lenton.2008}. In such cases, a subsystem may lose stability and pass from its current stable state to a new stable state, which might be quite different. A well-known example is the Atlantic Meridional Overturning Circulation (AMOC), a system of ocean currents that currently transports large amounts of heat from the tropics to Northwestern Europe, but at various times in the distant past has been ``switched off'' via such a tipping process \cite{Dijkstra_AMOC.2019,Lenton.2008}. A critical challenge for climate science is to identify whether such tipping is likely in the present state or in plausible future states of the climate system. Bearing in mind our incomplete observational knowledge of the real world, the answer to this question will necessarily be probabilistic. 

For low order deterministic conceptual models of such subsystems, and potentially also for complex climate models, these critical transitions are often caused by a change in the topological appearance of the system's phase portrait upon parameter variation, i.e.\ a \textit{bifurcation}. There is a well-developed and rich mathematical theory of these bifurcations \cite{Guckenheimer.1983,Kuznetsov.2004,Strogatz.2015,Wiggins.2003}. Classical bifurcation theory can be used to analyse the dynamical system's response upon a variation of system parameters. 

Now suppose we have a physically motivated model of such a subsystem that includes a range of linear and nonlinear feedbacks and consists of a parameter-dependent system of ordinary differential equations (ODEs)
\begin{align}
	\frac{dx}{dt} &= \dx = f(x,r), \label{eq:ODE}
\end{align}
where $f:\R^n\times\R^d \rightarrow \R^n$ is a vector field  (assumed sufficiently smooth here), and $r \in \R^d$ corresponds to the $d \in \N$ given model parameters.

In many practical climate models, the model parameters do not correspond directly to observable variables and so have to be estimated indirectly. The resulting parametric uncertainty can be taken into account by passing from the deterministic ODE system \eqref{eq:ODE} to a system of \textit{random} ODEs
\begin{align}
	\dx &= f(x,r(\omega)), \label{eq:rODE}
\end{align}
where $r(\omega) \in \R^d$ is a random vector corresponding to the $d$ uncertain model parameters on a fixed probability space $(\Omega,\AAA,P)$ with $\omega$ denoting the outcome from the sample space $\Omega$.
Within this framework for a climate model, two main questions arise:
\begin{enumerate}
	\item[(Q1)] How can the probability distribution of the model parameters be inferred?
	\item[(Q2)] How does this parametric uncertainty affect the tipping behavior of the system?
\end{enumerate}

\begin{figure*}[t]
	\centering
	\tikzstyle{arrow} = [thick,->,>=stealth, line width=2pt]
	\begin{tikzpicture}[scale=0.95]
		\node (knowledge) at (8,0) {\begin{minipage}{0.27\textwidth}\centering Knowledge of physical\\ processes + feedbacks\end{minipage}};
		
		\draw[draw=myblue2, fill=myblue2, rounded corners] (-3,-0.75) rectangle (3,0.75);
		\node (model) at (0,0) {\begin{minipage}{0.4\textwidth}\centering Climate model with\\ \textbf{uncertain parameters}\end{minipage}};
		\draw [arrow] (knowledge.west) -- ([xshift=0cm,yshift=0cm]model.east);
		
		\draw[draw=myblue1, fill=myblue1] ([xshift=-2.5cm,yshift=-1cm]model.south) ellipse (1.5cm and 0.5cm);
		\node at ([xshift=-2.5cm,yshift=-1cm]model.south) {\begin{minipage}{0.27\textwidth}\centering \textbf{Section \ref{subsec:expert}}\end{minipage}};

		\node (expert) at ([yshift=-2.5cm]knowledge.south) {\begin{minipage}{0.27\textwidth}\centering Expert knowledge \\ \& physical constraints\end{minipage}};
		
		\draw[draw=myblue2, fill=myblue2, rounded corners] ([xshift=-3cm,yshift=-3.25cm]model.south) rectangle ([xshift=3cm,yshift=-1.75cm]model.south);
		\node (ranges) at ([yshift=-2.5cm]model.south) {\begin{minipage}{0.4\textwidth}\centering Climate model with\\ plausible parameter ranges\end{minipage}};
		\draw [arrow] (expert.west) -- ([yshift=0cm]ranges.east);
		\draw [arrow] ([yshift=-0.6cm]model.center) -- ([yshift=1cm]ranges.center);

		\draw[draw=myblue1, fill=myblue1] ([xshift=-2.5cm,yshift=-1cm]ranges.south) ellipse (1.5cm and 0.5cm);
		\node at ([xshift=-2.5cm,yshift=-1cm]ranges.south) {\begin{minipage}{0.27\textwidth}\centering\textbf{Section \ref{subsec:BayesianInference}}\end{minipage}};
		
		\node (observations) at ([yshift=-2.5cm]expert.south) {\begin{minipage}{0.27\textwidth}\centering Simulations and/or \\ observational data\end{minipage}};
		
		\draw[draw=myblue2, fill=myblue2, rounded corners] ([xshift=-3cm,yshift=-3.25cm]ranges.south) rectangle ([xshift=3cm,yshift=-1.75cm]ranges.south);
		\node (paraDist) at ([yshift=-2.5cm]ranges.south) {\begin{minipage}{0.4\textwidth}\centering Climate model with\\ parameter probability distributions\end{minipage}};
		
		\draw [arrow] (observations.west) -- ([yshift=0cm]paraDist.east);
		\draw [arrow] ([yshift=-0.6cm]ranges.center) -- ([yshift=1cm]paraDist.center);
		
		\draw[draw=myblue1, fill=myblue1] ([xshift=-2.5cm,yshift=-1cm]paraDist.south) ellipse (1.5cm and 0.5cm);
		\node at ([xshift=-2.5cm,yshift=-1cm]paraDist.south) {\begin{minipage}{0.27\textwidth}\centering\textbf{Section \ref{sec:probTippingDiagram}}\end{minipage}};
		
		\draw[draw=black, fill=myblue1, rounded corners] ([xshift=-3cm,yshift=-6.25cm]paraDist.south) rectangle ([xshift=10cm,yshift=-1.75cm]paraDist.south);
		\node (outcome) at ([yshift=-2.5cm]paraDist.south) {};
		
		\draw [arrow] ([yshift=-0.6cm]paraDist.center) -- ([yshift=1cm]outcome.center);

		\node (hist) at ([xshift=-0.5cm,yshift=-4cm]paraDist.south) {\includegraphics[width=0.25\textwidth]{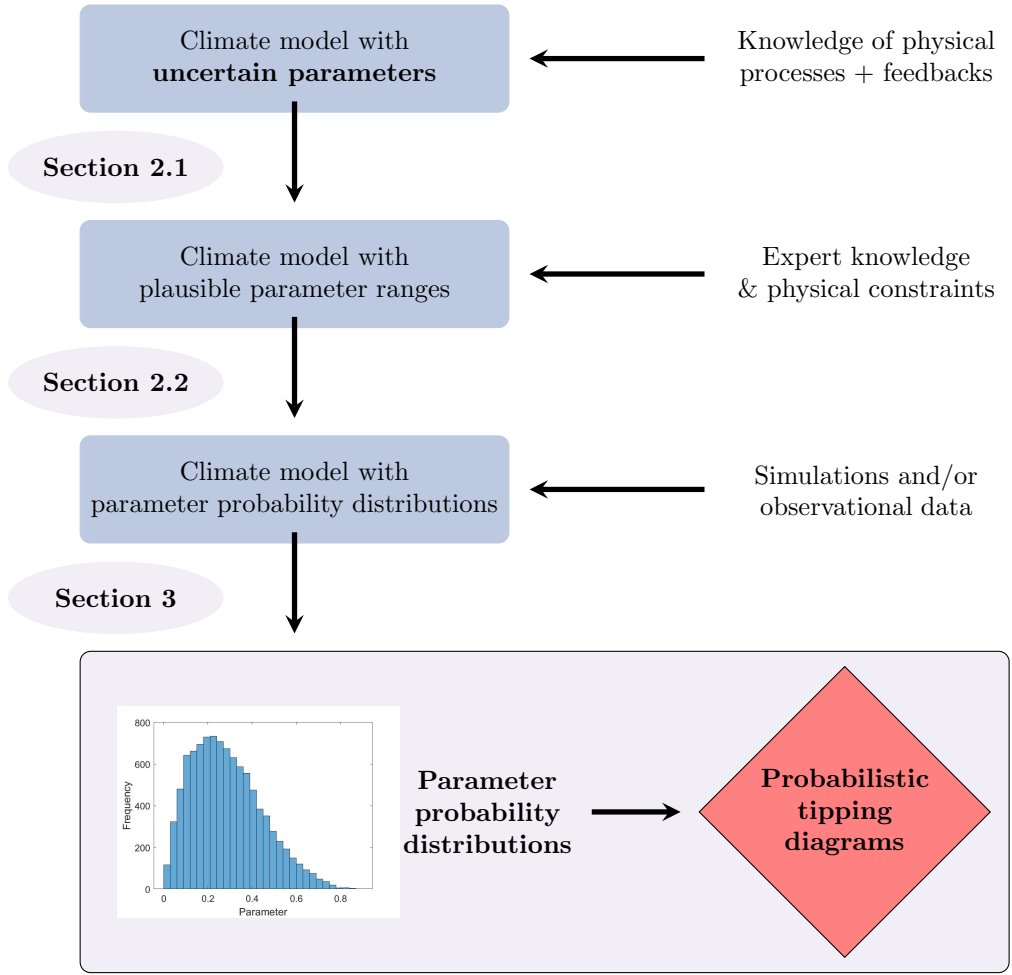}};
		\node (pdf) at ([xshift=3.2cm,yshift=0cm]hist.center) {\begin{minipage}{0.2\textwidth}\centering \textbf{Parameter probability distributions}\end{minipage}};
		\node[diamond, draw, fill=red!50] (diagram) at ([xshift=5cm,yshift=0cm]pdf.center) {\begin{minipage}{0.15\textwidth}\centering \textbf{Proba\-bilistic tipping diagrams}\end{minipage}};
		\draw [arrow] ([xshift=-0.25cm]pdf.east) -- ([xshift=-0.25cm]diagram.west);
	\end{tikzpicture}
	\caption{Outline of the steps described in this paper. Given a climate model, we provide prior ranges for uncertain model parameters and present a Bayesian inference method to derive corresponding parameter probability distributions. For visualization, we construct corresponding probabilistic tipping diagrams.
	}
	\label{fig:bigPicture}
\end{figure*}

With respect to the first question, physical constraints on the model parameters (sometimes involving an element of expert judgement), and  available observed data on the phenomenon described by the system (the state variable $x$) provide two key sources of information. This mathematically leads to an inverse problem of parameter fitting, which belongs to the huge research area of data assimilation and inverse problems. 
 
Many approaches to the parameter estimation problem have been investigated \cite{Jazwinski.1970,Stuart.2015,Stuart.2010,Sullivan.2015}. The use of Bayesian inference methods has become quite widespread in many areas \cite{Stuart.2015}, including climate science (see e.g.\ \cite{Huerta.2008}), and forms the basis of some climate change assessments using state-of-the art climate models \cite{Harris.2013,Tebaldi.2005}. 
As described in \cite{Green.2015}, one part of Bayesian system identification is concerned with model selection. However, here we start from a fixed given model and focus on the second part of the Bayesian system identification, namely the parameter inference as formulated in question (Q1). We illustrate our approach using a widely-studied model of AMOC tipping.

Only very recently, the link between parametric uncertainty and bifurcations has emerged as a sub-area within the theory of nonlinear dynamics~\cite{BredenKuehn1,HurthKuehn,KuehnDelayUQ,KuehnLux}. This link is important since parametric uncertainty leads to uncertainty in the location of tipping points and the shape of the bifurcation curve. An apparently small uncertainty in model parameters may lead  to a large uncertainty in tipping behaviour. Hence, a tipping point might be much closer to the current system state than expected from the 'best estimate' model parameters, or vice versa \cite{Lenton.2009}.

To develop a deeper understanding of the effects of \textit{parametric uncertainty} on tipping point dynamics, a holistic approach unifying aspects of the areas of bifurcation theory and data assimilation is needed. From the viewpoint of various applications, there is a clear need for additional technical tools. For example, in \cite{Tartaruga.2015}, the authors address the integration of parametric uncertainty into bifurcation diagrams in terms of confidence bounds and demonstrate their methodology in the context of a landing gear system. In \cite{Ring.2014}, polynomial chaos based approaches have been developed to generate probabilistic fixed point stability diagrams in the context of heart rhythm dynamics for given probability distributions. In \cite{Wang.2017}, the authors analyse how uncertainty in aerodynamic and structure parameters affects the bifurcation position. 

Here, we address the link between parametric uncertainty and tipping dynamics by presenting a framework to generate probabilistic tipping diagrams providing further visual insights on the sensitivity of the tipping behavior to the parametric uncertainty. Besides a probabilistic visualization of the classical bifurcation diagram, we come up with new visual aids to assess the risk of tipping.
In our work, we use a Bayesian approach exploiting the observed evolution of the system to provide more tightly constrained parameter distributions than are possible from broad physical constraints alone. We find that the Bayesian step provides substantially tighter constraints on the location of the tipping points.

The framework is sketched in Figure \ref{fig:bigPicture}. We use the Stommel-Cessi model \cite{Cessi.1994} for the AMOC to illustrate our methodology. Note however that our approach is not limited to this particular climate model. It might also be used for systems of random ODEs describing other elements of Earth System dynamics, or in other scientific disciplines such as neuroscience and epidemiology, where parametric uncertainty is very common. 

In our approach, no intermediate approximation step via an emulator or surrogate model has been undertaken. Hence, no model approximation error is induced. However, the incorporation of a surrogate model as in \cite{Ring.2014,Tartaruga.2015} could be relevant in future work with large general circulation models instead of conceptual box models \cite{Harris.2013}.

In Section \ref{sec:problemSetup}, we present our benchmark problem, the Stommel-Cessi box model \cite{Cessi.1994}, and its known deterministic tipping behavior.
We pre\-sent physically-constrained prior ranges for mod\-el parameters and present a Bayesian inference\\ method to assign corresponding probability distributions based on these physical constraints and available data.
Building upon the derived probability distribution, in Section \ref{sec:probTippingDiagram}, we construct corresponding probabilistic tipping diagrams to illustrate how the parametric uncertainty leads to uncertainty in tipping behaviour. We conclude this letter with a presentation of further applications and a short outlook.

\section{Problem setup} \label{sec:problemSetup}

To illustrate our methodology, we consider here the Stommel Cessi box model \cite{Cessi.1994}, which is a simplified description of the AMOC based on two boxes, an equatorial (e) and a polar (p) one. Temperature differences $\Delta T = T_e - T_p$ and salinity differences $\Delta S = S_e - S_p$ may arise due to an atmospheric freshwater flux $p(t)$ leading to a density gradient, in turn enabling a circulation between the boxes.

As described in \cite[Sec.\ 6.2.1]{Gentz.2006}, the system can be reduced to an attracting invariant manifold and after rescaling one obtains the reduced Stommel-Cessi dynamics as
\begin{align}
	\dot{x} &= \mu - x\left(1+\eta^2\left(1-x\right)^2\right) \label{eq:redStommelCessi}
\end{align}
with $x=\nicefrac{\alpha_S\Delta S}{\alpha_T \theta}$ corresponding to a dimensionless salinity difference, where $\alpha_S$ and $\alpha_T$ are saline contraction and thermal expansion coefficients and $\theta$ is the reference value for $\Delta T$. Note that small $x$ corresponds to the current strong AMOC state, while large $x$ corresponds to a weak or reversed AMOC.
More\-over, $\mu$ is proportional to the atmospheric freshwater flux, and 
\begin{equation}
\eta^2=t_d/t_a
\label{eq:etasq}
\end{equation}
is the ratio of the diffusive timescale $t_d$ to the advective timescale $t_a$.
The steady states are given by the manifold
\begin{align}
	C_0 = \left\{(x,\mu) \in (\mathbb{R}_0^+)^2 \ |\ \mu =x\left(1+\eta^2\left(1-x\right)^2\right) \right\} \label{eq:critManifold}
\end{align}
Its precise shape depends crucially on the parameter $\eta^2$. Although, for all $\eta^2$, the points $(\mu,x)=(0,0)$ and $(\mu,x)=(1,1)$ are equilibria, the system's dynamics show important qualitative differences for different values of $\eta^2$.

\textbf{Parameter regime: $\eta^2>3$}

\begin{figure}[h]
	\centering
	\begin{overpic}[width=0.5\textwidth]{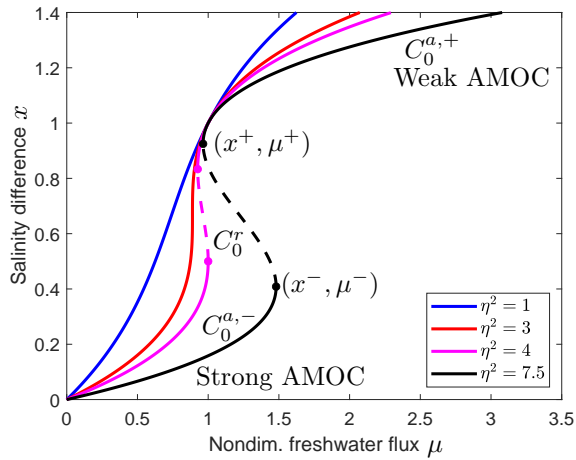}
	\put(70,70){$C_0^{a,+}$}
	\put(36,53.5){$\left(x^+,\mu^+\right)$}
	\put(36,36){$C_0^{r}$}
	\put(34,21){$C_0^{a,-}$}
	\put(48,28.5){$\left(x^-,\mu^-\right)$}
	\put(68,65){Weak AMOC}
	\put(33,12){Strong AMOC}
	\put(74,0){$\mu$}
	\put(1,59){\rotatebox{90}{$x$}}
	\end{overpic}
	
	\caption{The shape of the manifold \eqref{eq:critManifold} depends crucially on the value of the parameter $\eta^2$. In the parameter regime $\eta^2<3$, there are no tipping points whereas for $\eta^2>3$, the bifurcation curve is S-shaped and admits two fold points (see magenta respectively black dots).
	}
	
	\label{fig:redStommelCessi_critManifold}
\end{figure}

For $\eta^2>3$, $C_0$ has the classical S-shape. This means that there are two fold bifurcations linked to a variation in the bifurcation parameter $\mu$ with corresponding two fold points
\begin{subequations} 	\label{eq:foldPoints}
	\begin{align}
		\left(x^-,\mu^-\right) &= \left(\frac{2}{3} - A, \frac{2}{3} + \frac{2\eta^2}{27} + A\left(\frac{2\eta^2}{9} - \frac{2}{3}\right)\right), \label{eq:lowFoldPoint}\\
		(x^+,\mu^+) &= \left(\frac{2}{3} + A, \frac{2}{3} + \frac{2\eta^2}{27} - A\left(\frac{2\eta^2}{9} - \frac{2}{3}\right)\right), \label{eq:UpperFoldPoint}
	\end{align}
\end{subequations}
where $A=\sqrt{\nicefrac{1}{9}-\nicefrac{1}{(3\eta^2)}}$. The tipping points $(x^-,\mu^-)$ and $(x^+,\mu^+)$ split $C_0$ in an upper and lower attracting branch of fixed points $C_0^{a,-} = C_0 \cap \{x<x^-\}$ and $C_0^{a,+} = C_0 \cap \{x>x^+\}$ (upper and lower solid magenta and black lines in Figure \ref{fig:redStommelCessi_critManifold}). These correspond to stable weak and stable strong AMOC states. The two attracting branches are connected by a repelling branch $C_0^{r} = C_0 \cap \{x^-<x<x^+\}$ (dashed magenta and black line in Figure \ref{fig:redStommelCessi_critManifold}). A tipping of the AMOC from its current strong stable state to a new weak stable state might occur if $\mu$ were increased beyond $\mu^-$. This regime comes with a hysteresis behavior of the AMOC, which means that reversing the bifurcation parameter $\mu$ (proportional to nondimensional freshwater flux) to its value before the tipping occurred would not be enough. The freshwater flux would have to be further decreased below $\mu^+$ to get the former stable state back again.

\textbf{Parameter regime: $\eta^2<3$}

At $\eta^2=3$ (red solid line in Figure \ref{fig:redStommelCessi_critManifold}), we face a cusp bifurcation point. This means that, between the two attracting branches $C_0^{a,-}$ and $C_0^{a,+}$ of the double fold bifurcation, the connecting repelling branch $C_0^{r}$ ceases to exist.
Hence, for $\eta^2<3$, there are no longer values of $\mu$ that would allow for bistability of the system.

The true value of $\eta^2$ is unknown. Figure \ref{fig:redStommelCessi_critManifold} already reveals that it makes a major difference in the curvature of the manifold $C_0$ \eqref{eq:critManifold}, which value of $\eta^2$ we deal with.
It shows that a misspecification of the true value of $\eta^2$ can crucially change the overall qualitative behavior of \eqref{eq:redStommelCessi}. Therefore, a quantification of the effect of the uncertainty in $\eta^2$ is needed. First, we determine parameter ranges for $\eta^2$ based on physical constraints. In a second step, we infer a probability distribution of the parameter using a Bayesian inference technique. The bifurcation parameter $\mu$ is kept deterministic in our study.

\subsection{Physical constraints on model parameters} \label{subsec:expert}

The parameter $\eta^2$ plays the major role in our probabilistic tipping analysis. Different values for $\eta^2$ are used in the literature (e.g.\ $\eta^2=6.2$ and $\eta^2=6.25$ in \cite[Sec.\ 10.5]{Dijkstra.2013}) and $\eta^2=7.5$ in \cite{Gentz.2006,Cessi.1994}). Our aim is to derive a physically reasonable prior range for $\eta^2$ in \eqref{eq:redStommelCessi} before refining it using Bayesian inference.

Physically, $\eta^2$ represents the ratio of the advective timescale of salinity anomalies by the AMOC itself, to the exchange timescale due to advection and mixing by wind-driven gyres and eddies (represented by a diffusive timescale in the box model). Estimating appropriate values of these timescales is not straightforward because the boxes of the underlying two-box model \cite{Cessi.1994} cannot be directly associated with a closed representation of physical regions of the ocean. We therefore estimate the parameters using a slightly more complex box model \cite{Wood.2019}, which is a closed representation of the global ocean and so can be more directly associated with observations or with complex climate models. 

In \cite[Table 1]{Wood.2019}, AMOC box model parameters have been calibrated to seven General Circulation Model (GCM) states representing a \\ range of models of different generations. One of the model states used  is constrained to recent observations of temperature and salinity, while others represent climate states with different levels of atmospheric carbon dioxide between 1 and 4 times pre-industrial values. Across the seven model states, the gyre exchange rate ($K_N$) varies between $5.4$ and $20.9$ Sv and different values for the volume of the North Atlantic box ($V_N$) are given. This leads to a range for the diffusive timescale $t_d$ between $79$ and $662$ years. Based on the values for $V_N$ in \cite[Table 1]{Wood.2019} and assuming a MOC transport of $18$ Sv (relatively well constrained by observations \cite{Rayner.2011}), we obtain the range of $[57,92]$ years for the advective timescale $t_a$. On taking $t_d \in [60,700]$ years, we finally obtain from (\ref{eq:etasq}) the range for our parameter of interest, i.e.\ 
\begin{align}
	\eta^2 \in [0.6, 12.3]. \label{eq:etaP2range}
\end{align}
Note that an alternative, narrower physical range for $\eta^2$ can be obtained if we only consider the four GCM states that correspond to pre-indus\-trial or present day $CO_2$ levels. The ranges for $t_a$ and $t_d$ get narrower, leading to
\begin{align}
	\eta^2 \in [0.7,3.9]. \label{eq:etaP2range_narrow}
\end{align}
If we have access to time series data on the salinity difference or the AMOC strength, the physical constraint \eqref{eq:etaP2range} (or \eqref{eq:etaP2range_narrow}) can be used as \textit{prior information} for the Bayesian inference and we can
assign a probability distribution to $\eta^2$.

\subsection{Bayesian inference of model parameters} \label{subsec:BayesianInference}

Here, we perform Bayesian inference on the mod\-el parameter $\eta^2$ in \eqref{eq:redStommelCessi}. We use the MATLAB-based software UQLab \cite{Sudret.2014}, version 1.3.0. There\-in, a Markov Chain Monte Carlo (MCMC) approach is provided \cite{UQdoc13113}. For more information on MCMC methods, we refer the reader to \cite{Brooks.2011,Stuart.2010,Tierney.1994}. UQLab offers several algorithms within the MCMC framework. Here, we use the affine invariant ensemble algorithm (AIES) \cite{Goodman.2010,UQdoc13113} with 400 steps and 100 chains. Our setup is based upon the UQLab example\footnote{\url{https://www.uqlab.com/inversion-predator-prey}} of parameter estimation in a predator prey model.

\subsubsection{Synthetic data generation} \label{subsubsec:synthData}
Our forward model is given by \eqref{eq:redStommelCessi}. Since we do not have suitably long real data on the salinity differences or AMOC  available, we generate synthetic data to illustrate the approach. Note that the synthetic data could be replaced by appropriately scaled real salinity or AMOC data, or palaeo-proxy data, if available. For the generated 'truth' time series of salinity differences, we use a value of $\eta^2=4$ and simulate \eqref{eq:redStommelCessi} forward in time by using the MATLAB ODE solver \textit{ode45}\footnote{\url{https://de.mathworks.com/help/matlab/ref/ode45.html}, last checked: August 12, 2021} to showcase the methodology sketched in Figure \ref{fig:bigPicture}. We use a value of $\mu=0.85$ proportional to the nondimensional freshwater flux (equivalent to a fresh water flux of order 0.5 Sv into the North Atlantic basin), an initial salinity difference of $x_0=0.4$, a step size of $\Delta t=10^{-3}$, and a final time $T=5$ (corresponding to a dimensional integration timeseries length of around 1400 years). For the Bayesian inference, we just use every $100$th point of the time series (corresponding to sampling every 30 years) and put normally distributed noise with mean zero and a standard deviation of $0.3$ on the data points except for the first one (corresponding to a physical standard deviation of order 0.13 psu). 

\subsubsection{Prior and discrepancy model}
We specify a uniform prior over the physical parameter range \eqref{eq:etaP2range}. For the inference of the discrepancy variance, we use a lognormal prior with mean $-1$ and standard deviation $1$. The discrepancy is assumed to be Gaussian.

\begin{figure}[h]
	\centering
	\begin{overpic}[width=0.5\textwidth]{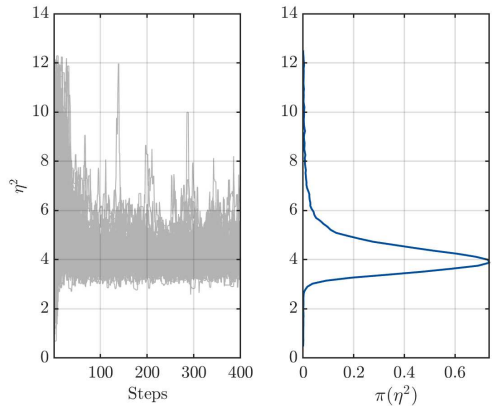}
	\end{overpic}
	
	\caption{Bayesian inference via UQLab \cite{UQdoc13113} with uniform prior $\UU(0.6,12.3)$: MCMC paths and posterior distribution of $\eta^2$ (mean: 4.1235, std: 0.5864).}
	
	\label{fig:MCMCpaths_y0_1K05_T5_etaP2_3K5_p0K9}
\end{figure}

The method works very well and the convergence of the Markov Chain Monte Carlo paths can be observed in Figure \ref{fig:MCMCpaths_y0_1K05_T5_etaP2_3K5_p0K9}. Note further that the posterior mean of $4.1235$ is very close to the true value of $\eta^2=4$ and the posterior standard deviation is $0.5864$.

\textbf{Robustness against prior information}

Our result is rather robust against variations in prior information. For a uniform prior $\eta^2\sim\UU(0,5)$, the posterior mean is $3.9815$ and the posterior standard deviation amounts to $0.4295$. For a Gaussian prior $\eta^2\sim\NN(6.45,4)$ ($\eta^2\sim\NN(4,0.25)$), where the first entry is the mean and the second entry the variance, the posterior mean is $4.3513$ ($4.0005$) and the posterior standard deviation (std) amounts to $0.6956$ ($0.3542$).

\textbf{Robustness against noise level}

If the noise in the data is too large relative to the values of the dynamic itself, the estimation procedure might not produce reliable results. By increasing the Gaussian noise standard deviation from $0.3$ to $0.5$, the posterior mean changes to $4.7042$ and the posterior standard deviation rises to $1.5139$. For a noise intensity of $1$, the posterior standard deviation further increases to $2.6745$ and the mean of $6.0544$ is rather far from the true value of $4$ but at least the mode of the posterior distribution is still close to the true value.

\textbf{Closeness of initial value to equilibrium value}

Moreover, the inference method seems to be capable of dealing with initial values of the time series close to an equilibrium point. For example, the choice of the initial value $x_0=0.3$ being very close to the equilibrium of $0.2729$ still recovers a posterior mean of $4.1593$ with a posterior std of $0.6674$ being only slightly larger than the posterior std of $0.5864$ for an initial value $x_0=0.4$. This is important because it is expected that any recent (holocene) AMOC proxy timeseries would be close to the strong AMOC equilibrium throughout its length; yet even variations close to one equilibrium state are giving information that constrains the tipping point.

\textbf{Robustness against discrepancy assumptions}

As a robustness check for the inference of the discrepancy variance, we replace the lognormal prior with mean $-1$ and standard deviation $1$ by a uniform prior over $[0,1]$ on the discrepancy variance. The model discrepancy is still assumed to be Gaussian. The mean and standard deviation of the posterior distribution of $\eta^2$ change only slightly (mean: 4.1462 and std: 0.6056).

\begin{figure}[h]
	\centering
	\begin{overpic}[width=0.5\textwidth]{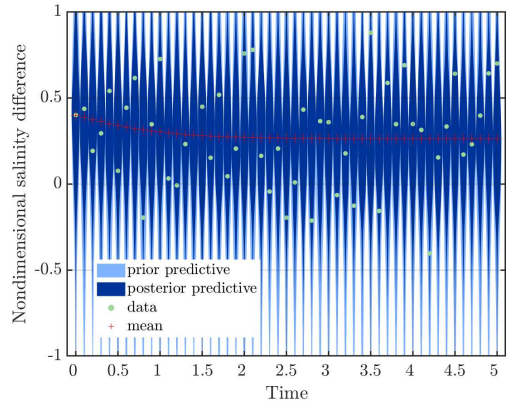}
	\end{overpic}
	
	\caption{Prior and posterior predictive distribution. The green synthetic data points are more narrowly comprised by the posterior predictive (dark blue) indicating a qualitative improvement within the inference process for $\eta^2$.}
	
	\label{fig:postPred_y0_1K05_T5_etaP2_3K5_p0K9}
\end{figure}

\bigskip
Figure \ref{fig:postPred_y0_1K05_T5_etaP2_3K5_p0K9} shows that the posterior predictive distribution gives a more narrow prediction of the green synthetic data points than the prior predictive distribution inidcating a qualitative improvement of the parameter estimate over the inference process.

\section{Probabilistic tipping diagrams} \label{sec:probTippingDiagram}
The prior physical ranges from Section \ref{subsec:expert} and the probability distribution obtained by the \\ Bayesian inference method in Section \ref{subsec:BayesianInference} can now be used to generate probabilistic tipping diagrams. These are intended to visually support the risk assessment of a possible tipping event.

\subsection{Tipping diagrams based on \\ physical parameter ranges}
\begin{figure}[h]
	\centering
	\begin{overpic}[width=0.5\textwidth]{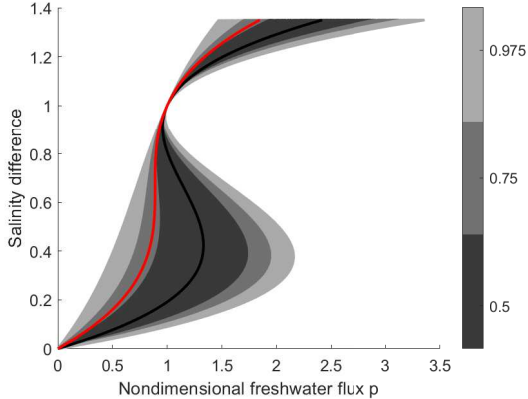}
	\end{overpic}

	\caption{How uncertainty in $\eta^2$ affects the shape of the manifold $C_0$ for $\eta^2 \sim \UU(0.6,12.3)$. The red line in this and subsequent diagrams corresponds to the $\eta^2 = 3$ case.}
	
	\label{fig:critMani_StommelCessi_uniform}
\end{figure}

\begin{figure}[h!]
	\centering
	\begin{overpic}[width=0.5\textwidth]{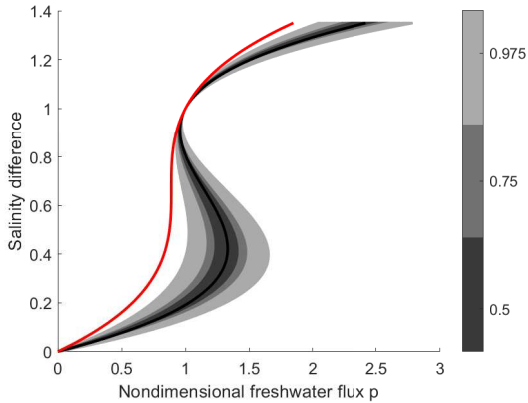}
	\end{overpic}
	
	\caption{How uncertainty in $\eta^2$ affects the shape of $C_0$ for $\eta^2 \sim \NN_{0.6}^{12.3}(6.45,1)$.}
	
	\label{fig:critMani_StommelCessi_truncNormal}
\end{figure}

We use the prior parameter range \eqref{eq:etaP2range} for the parameter $\eta^2$. In Figure \ref{fig:critMani_StommelCessi_uniform}, we plot the manifold $C_0$ of steady states \eqref{eq:critManifold} for $\eta^2 \sim \UU(0.6,12.3)$, i.e.\ $\eta^2$ is uniformly distributed on the interval $[0.6,12.3]$.
The grey scale shows the mass levels indicating the percentages of realizations that are covered by the corresponding area. These are calculated numerically based on $M=10^3$ realizations. The red solid line in Figures \ref{fig:critMani_StommelCessi_uniform}, \ref{fig:critMani_StommelCessi_truncNormal} and subsequent figures indicates the bifurcation curve for the cusp value $\eta^2=3$. The black solid line corresponds to the mean $\eta^2=6.45$ and obeys the characteristic S-shape described in Section \ref{sec:problemSetup}. We observe that, due to the nonlinearity, the uniform values of $\eta^2$ are squeezed and expanded depending on the salinity difference level $x$.

Furthermore, the prior assumption on the parameter distribution can have a major impact on the risk assessment for the system to undergo a tipping. If we assume a truncated normal distribution over the physical range \eqref{eq:etaP2range} for $\eta^2$ with mean $6.45$ and standard deviation of 1, i.\ e.\ $\eta^2 \sim \NN_{0.6}^{12.3}(6.45,1)$, we obtain a more narrow area for the bifurcation curves being comprised by the bistable regime $\eta^2>3$.
Whereas a value of $\eta^2=3$ is within the range where $75\%$ of the realizations lie for the uniform distribution over the given physical range (compare Figure \ref{fig:critMani_StommelCessi_uniform}), $\eta^2=3$ is not contained in $97.5\%$ of the realizations for a truncated normal distribution $\eta^2\sim \NN_{0.6}^{12.3}(6.45,1)$ (compare Figure \ref{fig:critMani_StommelCessi_truncNormal}).

\subsection{Tipping diagrams based on \\ Bayesian parameter inference}
\begin{figure}[h]
	\centering
	\begin{overpic}[width=0.5\textwidth]{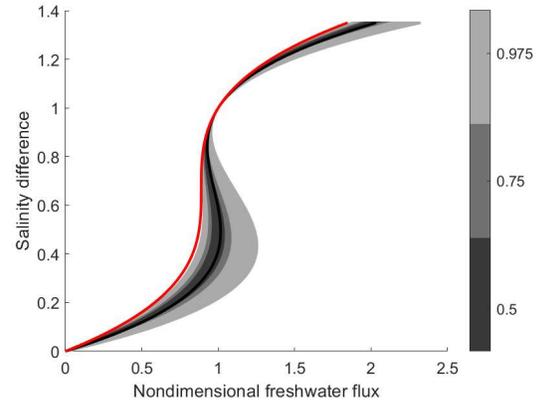}\end{overpic}
	
	\caption{Probabilistic bifurcation diagram for Stommel-Cessi model resulting from Bayesian inference on synthetic data from Section \ref{subsubsec:synthData} with uniform prior according to \eqref{eq:etaP2range}.}
	
	\label{fig:probBifCurves_y0_1K05_T5_etaP2_3K5_p0K9_noise0K05}
\end{figure}

Therefore, the aim is now to include more information on how the parameter values for $\eta^2$ are actually distributed. Building upon the physical parameter ranges, we can perform a Bayesian inference analysis as described in Section \ref{subsec:BayesianInference}.
This gives us an estimate of the probability distribution function (PDF) of our parameter $\eta^2$. Here, we use a sample of size $M=20,100$ drawn from the posterior distribution exported from our UQLab result. In the same manner as Figures \ref{fig:critMani_StommelCessi_uniform} and \ref{fig:critMani_StommelCessi_truncNormal}, Figure \ref{fig:probBifCurves_y0_1K05_T5_etaP2_3K5_p0K9_noise0K05} shows the sample-based probabilistic tipping diagram. The area in which the bifurcation curves lie is  far narrower than under the uniform distribution assumption and is based on the synthetic data from Section \ref{subsubsec:synthData}. In particular the range of fresh water flux $\mu$ for which a strong AMOC can be maintained is much narrower.  Note that, for a true value of $\eta^2=4$, no realizations in the regime without tipping points occur any longer. The Bayesian step has drastically reduced the dependence of the tipping point estimate on the somewhat arbitrary choice of prior distribution.

\subsubsection{Visualization of probabilistic tipping points}
\begin{figure}[h]
	\centering
	\begin{overpic}[width=0.5\textwidth]{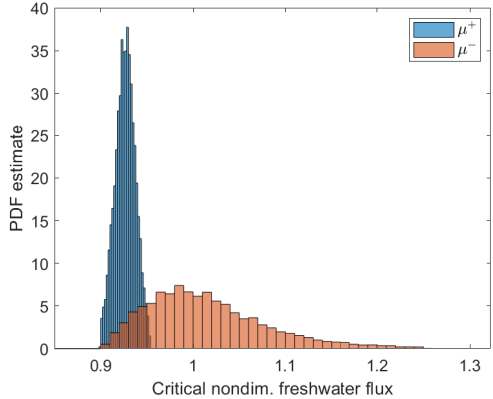}
	\end{overpic}
	
	\caption{Histogram of critical values of nondimensional freshwater flux for strong and weak AMOC tipping points $(x^-,\mu^-)$ and $(x^+,\mu^+)$ based on posterior sample of $\eta^2$.}
	
	\label{fig:hist2DcritFreshwater}
\end{figure}

\begin{figure}[h]
	\centering
	\begin{overpic}[width=0.5\textwidth]{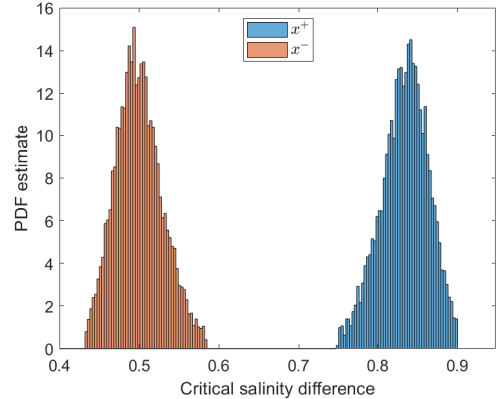}
	\end{overpic}
	
	\caption{Histogram of critical values of salinity difference for strong and weak AMOC tipping points $(x^-,\mu^-)$ and $(x^+,\mu^+)$ based on posterior sample of $\eta^2$.}
	
	\label{fig:hist2DcritSalDiff}
\end{figure}

The parameter uncertainty in $\eta^2$ does not only turn the bifurcation curves into random objects but also the location of the tipping points themselves becomes uncertain.
Histograms based on the posterior sample of $\eta^2$ for critical values of the nondimensional freshwater flux as well as critical values of the salinity difference are depicted in Figures \ref{fig:hist2DcritFreshwater} and \ref{fig:hist2DcritSalDiff}.

\begin{figure}[h]
	\centering
	\begin{overpic}[width=0.5\textwidth]{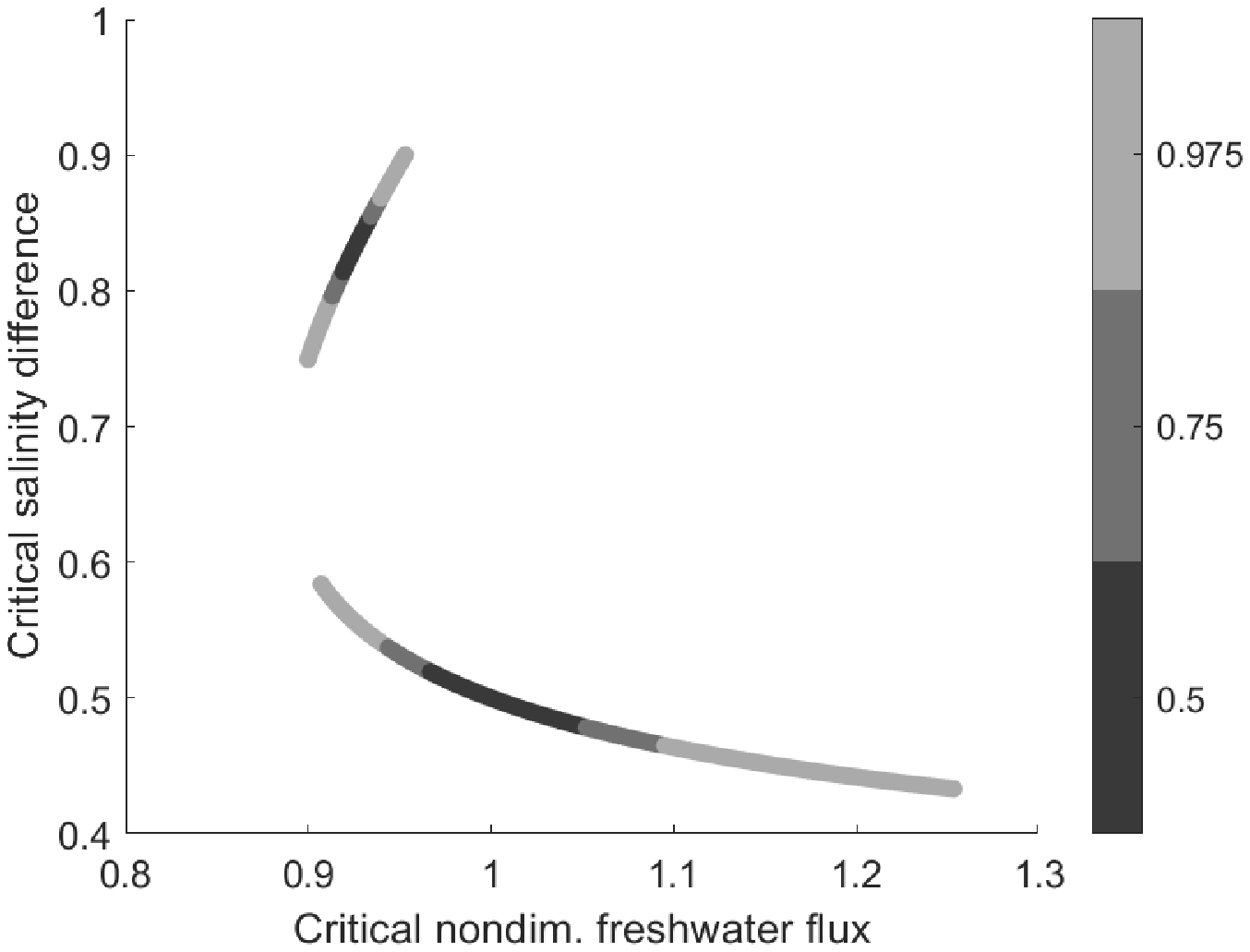}
	\end{overpic}
	
	\caption{Probabilistic location of strong and weak AMOC tipping points $(x^-,\mu^-)$ and $(x^+,\mu^+)$ based on posterior sample of $\eta^2$.}
	
	\label{fig:probLocationTPs}
\end{figure}

By putting together the critical values $\mu^-$ and $\mu^+$ for the nondimensional freshwater flux with the critical values $x^-$ and $x^+$ for the salinity difference, we obtain the probabilistic representation of the location of the weak and strong AMOC tipping points $(x^-,\mu^-)$ and $(x^+,\mu^+)$ in Figure \ref{fig:probLocationTPs}. The grey scale again indicates the ranges, in which the tipping points lie for the indicated masses of realizations of $\eta^2$. Figure \ref{fig:probLocationTPs} reveals that the location of the tipping points varies significantly for the sample values from the posterior distribution of $\eta^2$.

\subsubsection{Probabilsitic visualization of cusp bifurcation}
\begin{figure}[h]
	\centering
	\begin{overpic}[width=0.5\textwidth]{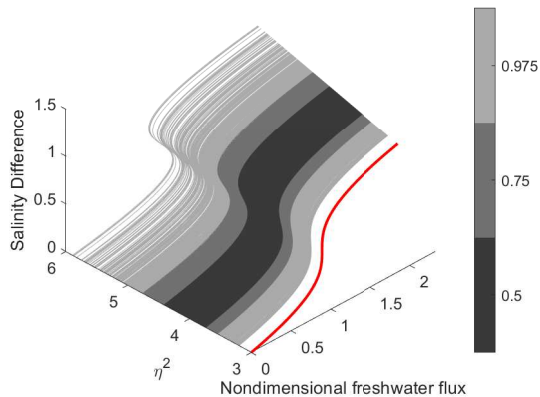}
	\end{overpic}
	
	\caption{Probabilistic fold curves in 3D.}
	
	\label{fig:probBifurcationCurves3D}
\end{figure}

By passing from a 2D representation as in Figure \ref{fig:probBifCurves_y0_1K05_T5_etaP2_3K5_p0K9_noise0K05} to a 3D representation and again using the grey scale color code as before, we provide a probabilistic visualization of the fold curves with corresponding values of $\eta^2$ (see Figure \ref{fig:probBifurcationCurves3D}). Observe again the occurrence of a cusp bifurcation at $\eta^2=3$: there is no value of the nondimensional freshwater flux for which bistability of the system occurs in front of the red curve ($\eta^2=3$) and the classical S-shape double fold curve can be observed behind the red curve.

\begin{figure}[h]
	\centering
	\begin{overpic}[width=0.5\textwidth]{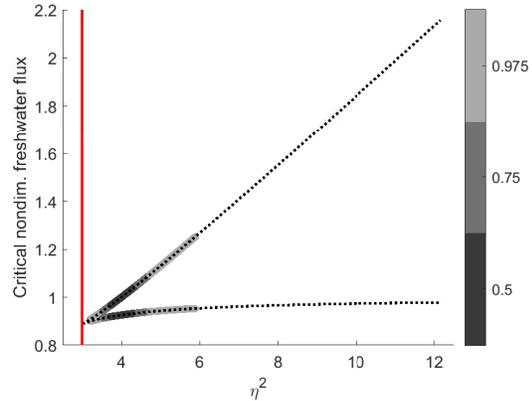}
	\end{overpic}
	
	\caption{Visualization of cusp bifurcation at $\eta^2=3$ for critical values of nondimensional freshwater flux corresponding to the posterior sample of $\eta^2$ (grey scale). The dotted line shows the range of bifurcation positions obtained when using the uniform prior distribution for $\eta^2$. }
	
	\label{fig:probCritFreshwater}
\end{figure}

The probabilsitic cusp bifurcation behavior can also be visualized by plotting realizations of the posterior distribution of the uncertain model parameter $\eta^2$ against the corresponding critical nondimensional freshwater flux values (see Figure \ref{fig:probCritFreshwater}). Again, the grey scale indicates the spread of the realizations of $\eta^2$ and the red line the cusp value of $\eta^2=3$. We added the black dotted line of critical nondimensional freshwater flux values according to the prior distribution of $\eta^2$. Note again that the range of fresh water flux $\mu$ for which a strong AMOC can be maintained is much narrower for the posterior sample.

\subsubsection{Holistic probabilistic tipping visualization}
\begin{figure*}[h]
	\centering
	\subfloat[\label{fig:probBifDiagramPlusTP}\ 3D-view]{\begin{overpic}[width=0.5\textwidth]{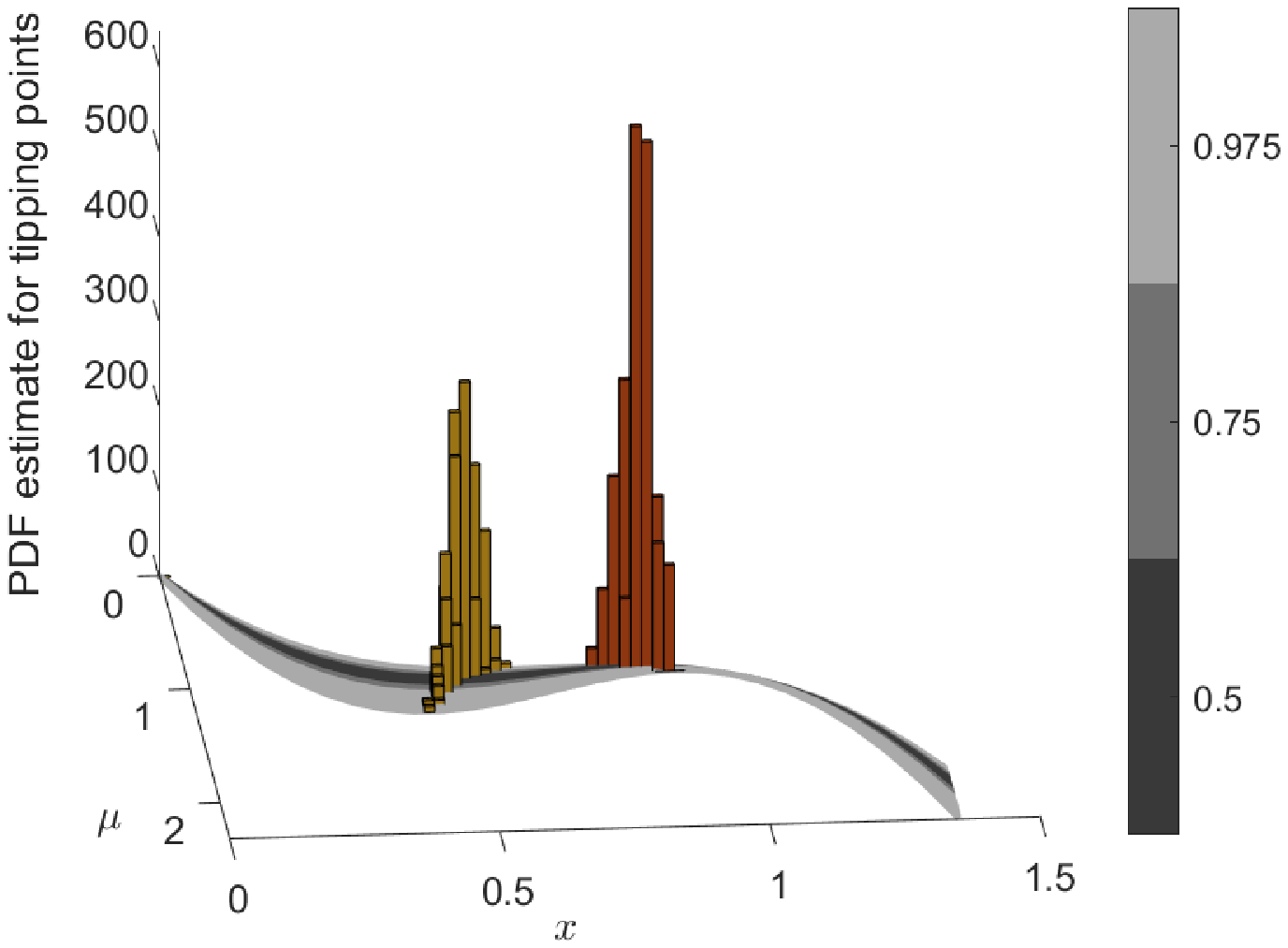}
	\end{overpic}}
	\subfloat[\label{fig:probBifDiagramPlusTP_topDown}\ Projection on $(x,\mu)$-plane]{\begin{overpic}[width=0.5\textwidth]{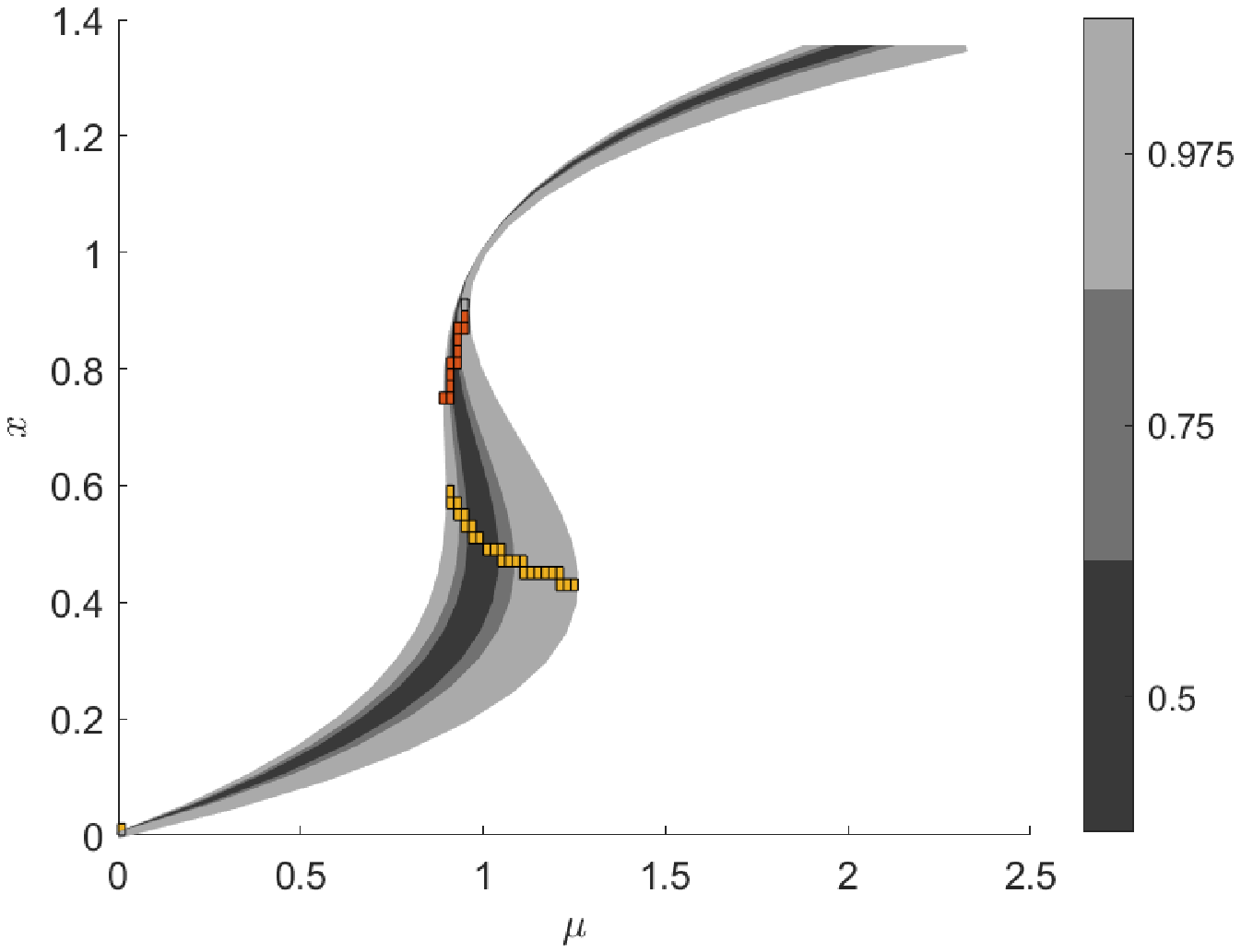}
	\end{overpic}}
	
	\caption{Probabilistic bifurcation diagram for Stommel-Cessi model resulting from Bayesian inference on synthetic data from Section \ref{subsubsec:synthData} with probabilistic tipping points locations.}
	
	\label{fig:probBifDiagramPlusTP_2views}
\end{figure*}

In a final visualization (Figure \ref{fig:probBifDiagramPlusTP_2views}), we collect insights that we have gained from previous diagrams. We combine the probabilistic bifurcation diagram from Figure \ref{fig:probBifCurves_y0_1K05_T5_etaP2_3K5_p0K9_noise0K05} with the probabilistic location of the tipping points shown in Figure \ref{fig:probLocationTPs}. A top-down view is shown in Figure \ref{fig:probBifDiagramPlusTP_topDown}. The grey scale color code is used for the probabilistic bifurcation curves as before and the probabilistic location of the tipping points is depicted thereon by using a histogram based on the posterior sample of $\eta^2$ in the third dimension (see Figure \ref{fig:probBifDiagramPlusTP}).

\section{Summary and outlook}

In this work, we have proposed new techniques to understand the influence of parameter uncertainty on tipping diagrams. We have described the required steps to obtain a probabilistic estimate of the model parameter(s) and shown that, given a suitable observed timeseries, a Bayesian approach can drastically and robustly narrow uncertainty in the location of the bifurcation point. We have proposed visualization techniques to understand the impact of parameter uncertainty on bifurcations. We have illustrated our results in the context of an AMOC box model. Of course, several natural extensions of our approach are possible, e.g., to larger parameter spaces, higher dimensional models, and more complex bifurcation phenomena. Other interesting future research directions are to develop automated and optimized numerical algorithms to generate probabilistic bifurcation diagrams for random differential equations, and to explore the use of real observational or proxy constraints. As increased emphasis is given to preparedness for high impact but low likelihood climate outcomes such as tipping points \cite{IPCC.2021}, the techniques we show here are likely to prove a valuable tool in climate risk assessment.

\bigskip
\textbf{Acknowledgment:}
The authors acknowledge support of the EU within the TiPES project funded by the European Union's Horizon 2020 research and innovation programme under grant agreement No.\ 820970.

\bigskip
\textbf{Data availability statement}
The data that support the findings of this study are available upon reasonable request from the authors.

\bibliographystyle{siam}
\bibliography{myLiterature}


\end{document}